\documentclass[12pt,epsf]{article}

\usepackage{amsmath,amssymb}
\usepackage{cite}
\usepackage{graphicx}

\DeclareMathOperator{\real}{Re}

\newcommand{\R}{\mathbb{R}}

\begin{document}
\begin{titlepage}
\begin{center}

 \vspace{-0.7in}

{\large \bf The Distributional Zeta-Function \\
\vspace{.06in} in\\
\vspace{.06in}
Disordered Field Theory}\\
\vspace{.5in}
{\large\em
B. F. Svaiter,\,\,\footnotemark[1]  N. F. Svaiter\,\,\footnotemark[2]}\\
\vspace{.08in}

Instituto de Matem\' atica Pura e Aplicada - IMPA \footnotemark[1] \\
Estrada Dona Castorina 110 Rio de Janeiro.
 RJ 22460-320, Brazil\\
\vspace{.06in}

Centro Brasileiro de Pesquisas F\'\i sicas - CBPF \footnotemark[2]\\
Rua Dr. Xavier Sigaud 150
Rio de Janeiro, RJ,22290-180, Brazil\\

\subsection*{\\Abstract}
\end{center}

In this paper we present a new mathematical rigorous technique for computing the average free
energy of a disordered system with quenched randomness, using the replicas.
The basic tool of this technique is
a distributional zeta-function, a complex function
whose derivative at the origin yields the average free energy of the system as the sum of two
contributions: the first one is
a series in which all the
integer moments of the partition function of the model contribute; the second one, which can not be
written as a series of the integer moments, can be made as small as desired.
This result supports the use of integer moments of the partition function, computed via replicas, for
expressing the average free energy of the system.
One advantage of the proposed formalism is that it does not require
the understanding of the properties
of the permutation group when
the number of replicas goes to zero.
Moreover, the symmetry
is broken using the saddle-point equations of the model.
As an application for the distributional zeta-function technique, we obtain the average free energy of the
disordered $\lambda\varphi^{4}$ model defined in
a $d$-dimensional Euclidean space.

\bigskip

\vspace{.06in}

\noindent
{\sc keywords:} disordered systems; average free-energy; replicas; distributional zeta-function.

\vspace{.06in}

\footnotetext[1]{e-mail:\,\,benar@impa.br}
\footnotetext[2]{e-mail:\,\,nfuxsvai@cbpf.br}
\vspace{.09in}
\noindent
PACS numbers: 05.20.-y,\,75.10.Nr

\end{titlepage}
\newpage\baselineskip .18in

\section{Introduction}
\quad

Disordered systems have been investigated for decades in statistical mechanics
\cite{binder,livro1,livro2,livro3,livro4}, condensed matter,
gravitational physics \cite{pe2,pe10,pe11,dis1,dis2,dis3} and even number theory \cite{jpa}.
The physics of spin glasses, disordered electronic systems and directed polymers in random media
are well known examples of such systems.
In statistical mechanics, the disorder may significantly affect critical behavior of different systems \cite{Harris}.
An example of a system with disorder is the random field Ising model \cite{larkin}, whose
hamiltonian is analogous to the one of the
classical Ising model, but allowing for a quenched random magnetic field.
This model can be analyzed using a continuous description, instead of
dealing with discrete elements. The continuous model is obtained in the limit of zero lattice spacing. In this case,
the hamiltonian of the discrete model can be replaced by an effective hamiltonian of the
Landau-Ginzburg model, where the order parameter is a continuous field coupled with
a quenched random field.
This continuous model has been extensively studied
using statistical field theory methods (e.g.,
 \cite{parisi1,mezard1, mezard2, dotsenko, orland,dotsenko2, dotsenko3,sherington1,sherington}).
The main problem of this disordered system is that a given realization of disorder is not invariant by
translation and the correlation functions are not similar to the ones of translationally invariant systems. Also,
in the presence of disorder, ground state configurations of the
scalar field are defined by a saddle-point equations and
the solutions of such equation depend on particular configurations of the random field.
Due to the presence of these local minima, it is very difficult to implement a perturbative approach in a straightforward way.

An established  procedure to circumvented the above discussed problems is to average
some thermodynamic extensive quantity with respect to the probability distribution of the disorder.
From a disordered system, where all the correlation
functions are not translational invariant, averaging on the disorder one obtains a
system where the correlation functions are translational invariant.
However, this procedure is not always straightforward.
For quenched disorder, one is mainly interested in
averaging the free energy over the disorder, which amounts to averaging the log of the partition function $Z$.
A well know technique for computing the average free energy is the replica method \cite{edwards} based in results discussed in Ref.\cite{polya}.
In this approach,
a new system defined by $k$ statistically independent replicas is introduced and the limit $k \to 0$ is taken.
In this process, one has to understand the behavior of the symmetry group $S_k$ as $k\to 0$.
Moreover, special procedures are sometimes required for obtaining sound physical results.
For example, the average free energy of the Sherrington-Kirkpatrick model \cite{sk}, which is the infinite-range version of the
Edwards-Anderson model, exhibits a (problematic)
\emph{negative entropy} at low temperatures, assuming a replica symmetric solution. 
The scheme of replica symmetry breaking was introduced to avoid this unphysical result. 
A step further was introduced by Parisi  with another replica
symmetry breaking solution, still using the replica formalism \cite{pa1,pa2,pa3},
by choosing a suitable 
ultrametric parametrization of
the replica matrix in the computation of the average free energy. See also the Ref. \cite{pa4}.
This scheme 
describes many stable states with ultrametric structure in the phase space, in which
the low-temperature phase consists of infinitely many pure
thermodynamic states.

Despite the success in the application of the replica method in disordered systems,
some authors consider that a mathematical rigorous derivation to support this procedure is still lacking
\cite{cri1,cri2,cri4,cri3}.
It is therefore natural
to ask whether there exists
a mathematically rigorous method, based on the use of replicas, for computing the
average free energy of systems with quenched disorder.
In Ref. \cite{sum}, Dotsenko considered an alternative approach where the summation of
all integer moments of the partition function is used
to evaluate the average free energy of the random energy model \cite{derrida2,derrida}. Also
a replica calculation using only the integer moments of the
partition function have been considered in Ref. \cite{virasoro}.
Some
alternative approaches to the replica method are the dynamical approach \cite{dynamical}
and the cavity method \cite{cavity}. In the Ref. \cite{livromez}, the main concepts and methods of the cavity method is discussed.

In this paper we define a new mathematical object associated with systems with quenched disorder,
namely, a complex function which, due to its similarities with zeta-functions, we will call distributional zeta-function.
This terminology is in light with a probabilistic approach, where we introduce a probability distribution to define this zeta-function.
We will show that the derivative of this function at the origin, which is well and uniquely defined, 
yields the
average free energy of the underlying system with quenched disorder.
Unlike the standard replica method, this procedure
does not require derivation with respect to the (integer) moments of the partition function.
Moreover,
in our derivation the average free energy of the random
$\lambda\varphi^{4}$ model is given by a series in which all the replicas contribute.
This method is closely related to the use of spectral zeta-functions
for
evaluating the determinant of differential operators of different systems in Euclidean field
theory \cite{seeley,ray,hawking,dowker,fulling}
and for computing 
the renormalized vacuum energy
of quantum field in the presence of boundaries. Although different global methods can be used to obtain the
Casimir energy of quantum fields, as for example an exponential cut-off or an analytic
regularization procedure  \cite{nami1,nami2,nami3}, the
spectral zeta-function method is powerful, elegant and widely used in quantum field theory \cite{elizalde}.

The organization of this paper is as follows.
In section II we discuss the replica field theory obtained from a quenched disordered
field linearly coupled to the Euclidean scalar
field.
In section III we
introduce the distributional zeta-function
and compute the derivative of the distributional zeta-function at the origin in order to obtain the
average free energy of the system with quenched disorder.
Conclusions are given in section IV. In the appendix we briefly review the use of the
spectral zeta-function method in Euclidean field theory.
We use $\hbar=c=k_{B}=1$.

\section{The replica field theory for the disordered scalar model}

\quad

Our aim is to use the Euclidean formalism of field theory to discuss classical statistical systems with disorder.
We study a field theory with fluctuations around the mean field solution.
Let us assume, for simplicity,  that
we have only a quenched disordered field linearly coupled to a Euclidean massive scalar
field. In this case,
one must obtain average values of extensive quantities.
An establish technique for
computing the average free energy is the replica method,
which we briefly describe.
The key point of this method is to compute integer
moments of the partition function $\mathbb{E}\,{Z^{k}}$ and to use such information to calculate
$\mathbb{E}\,{\ln Z}$.
The replica method consists in the following steps.
First, one constructs
the (integer) $k$-th power of the partition function
$Z^{\,k}=Z\times Z\times...\times Z$,
and interprets it as the partition function of a \emph{new} system, formed of $k$ statistically
independent copies of the original system.
Second, the expected value of the partition function's $k$-th power $Z_k=\mathbb{E}\, Z^k$ is computed by integrating over the disorder field
on the new model.
Finally,
the average free energy is obtained using the formula
 $\mathbb{E}\,{\ln Z}=
\lim_{k\rightarrow 0}\frac{Z_{k}-1}{k}$, where $Z_{k}$ for $0<k<1$ is derived from its values for $k$ integer.
Note that in $Z_k$, integration over the disorder field yields a system defined by
$k$ replicas which are no more statistically independent.
%
The average value of the free energy in the presence of the quenched disorder
is then obtained in the limit of a zero-component
field theory, taking the limit $k\rightarrow 0$.
Therefore, in this procedure it is necessary to understand the properties of the
permutation group $S_{k}$, when $k\,\rightarrow\,0$.

We start with the usual $\lambda\varphi^{4}$ Euclidean scalar field theory in the
presence of a disorder field $h(x)$.  In this setting, the functional integral
$Z(h)$ is defined by \cite{livron}:
\begin{equation}
Z(h)=\int [d\varphi]\,\, \exp\left(-S+ \int d^{d}x\,
h(x)\varphi(x)\right),
\label{8}
\end{equation}
where $S=S_{0}+S_{I}$  is the action that usually describes a massive scalar field with
the contributions $S_{0}$ and $S_{I}$ given, respectively, by
\begin{equation}
S_{0}(\varphi)=\int d^{d}x\, \left(\frac{1}{2}
(\partial\varphi)^{2}+\frac{1}{2}
m_{0}^{2}\,\varphi^{2}(x)\right),
\label{9}
\end{equation}
and
\begin{equation}
S_{I}(\varphi)= \int d^{d}x\,\frac{g_{0}}{4!} \,\varphi^{4}(x).
\label{10}
\end{equation}
The term $S_0$ is the free field action while $S_I$, a non-Gaussian contribution,
defines the
interacting part.
In Eq. (\ref{8}), $[d\varphi]$ is a functional measure,
formally given by $[d\varphi]=\prod_{x} d\varphi(x)$. The terms
$g_{0}$ and $m_{0}^{2}$ are, respectively, the bare coupling constant
and the bare mass square of the model.
The term $h(x)$ is a quenched random field. A normalization factor appears
in the definition of the gaussian functional integral,
${\cal{N}}=\left(\mbox{det}(-\Delta+m_{0}^{2})\right)^{\frac{1}{2}}$,
but in the following, as usual, we are absorbing this normalization factor in the
functional measure. The symbol $\Delta$ denotes the Laplacian in $\mathbb{R}^{d}$.

One comment is in order. In the above formal functional integral formulation there are two kinds of random  variables. The first one are
the Euclidean fields $\varphi_{i}(x)$. These fields describes generalized random  processes
with zero mean and covariance $G_{0}(x-y,m_{0})=\langle\,x|(-\Delta+m_{0}^{2})^{-1}|\,y\rangle$. The second one,
are the disordered field $h(x)$, in which the  
the absence of any gradient either makes them  statistically independent in every point of
the domain or, for fields that are not statistically independent for different points
of the domain, the two-point correlation function is not defined in terms of gradients.
We call these as uncorrelated and correlated disordered fields, respectively.
We assume that the random variables characterizing the disorder exhibit no long-range correlations, therefore the
probability distribution is written as
\begin{equation}
P(h)=p_{0}\,\exp\Biggl(-\frac{1}{2\,\sigma}\int\,d^{d}x(h(x))^{2}\Biggr).
\label{dis2}
\end{equation}
The quantity $\sigma$ is a small positive parameter associated with the disorder and $p_{0}$ is
a normalization constant.
In this case we have a delta correlated random field, with two-point correlation function
$\mathbb{E}({h(x)h(y)})
=\sigma\delta^{d}(x-y)$.
In the next paragraph, we will obtain the replica partition function,
$\mathbb{E}\,{Z^{\,k}}$, for this
disordered
model.

Let us construct $Z^{\,k}=Z\times Z\times...\times Z$.
We interpret $Z^{\,k}$ as the partition function of a new system, formed from $k$ statistically
independent copies of the original system. We have that the partition function  $Z^{\,k}$ of the replicated system is
\begin{equation}
(Z(h))^{k}=\int\,\prod_{i=1}^{k}[d\varphi_{i}]\,\exp\biggl(-\sum_{i=1}^{k}S(\varphi_{i},h)\biggr).
\label{u1}
\end{equation}
Let use the fact that $h$ is a random variable with probability distribution $d\rho[h]=d[h] P(h)$.
Averaging the free energy over the disorder
we obtain that the replica partition function
$Z_{k}$ is given by
\begin{equation}
Z_{\,k}=\int\,\prod_{i=1}^{k}[d\varphi_{i}]\,\exp\biggl(-S_{eff}(\varphi_{i},k)\biggr),
\label{aa11}
\end{equation}
where the effective action $S_{eff}(\varphi_{i},k)$ is
\begin{equation}
S_{eff}(\varphi_{i},k)=\frac{1}{2}\sum_{i,j=1}^{k}
\int d^{d}x\int d^{d}y\,\,\varphi_{i}(x)
D_{ij}(x-y)
\varphi_{j}(y)+\frac{\lambda}{4!}\sum_{i=1}^{k}\int\,d^{d}x\, \varphi_{i}^{4}(x).
\label{aa12}
\end{equation}
In the above equation $D_{ij}(x-y)=D_{ij}(m_{0},\sigma;x-y)$ where
\begin{equation}
D_{ij}(m_{0},\sigma;x-y)=\biggl(\delta_{ij}
(-\Delta+m_{0}^{2})-\sigma\biggr)\delta^{d}(x-y).
\label{aa124}
\end{equation}
The saddle-point equations of this model are given by
\begin{equation}
\Bigl(-\Delta\,+m_{0}^{2}\Bigr)
\varphi_{i}(x)
+\frac{\lambda}{3!}\varphi^{3}_{i}(x)=\sigma\sum_{j=1}^{k}\varphi_{j}(x).
\label{sp2}
\end{equation}
Let us use a replica symmetric ansatz, i.e., suppose that $\varphi_{i}(x)=\varphi(x)$. For equal
replicas, the saddle-point equations becomes a single equation given by
\begin{equation}
\Bigl(-\Delta\,+m_{0}^{2}-k\sigma\Bigr)
\varphi(x)
+\frac{\lambda}{3!}\varphi^{3}(x)=0.
\label{sp}
\end{equation}
In this case we must have $m_{0}^{2}-k\sigma\,>0$ to have a physical theory.
The simplest case of identical replicas leads to the replica symmetry breaking situation.

Let us discuss first the perturbative expansion of the replica field theory.
Starting from a gaussian theory, let us define $Z_{0\,k}(J_{i})$ where we introduce external sources $J_{i}(x)$ which are smooth
functions that we introduce to generate the correlation functions of
the theory by functional derivatives. We have
\begin{equation}
Z_{0\,k}(J_{i})
=\int\prod_{i=1}^{k}[d\varphi_{i}]\exp{\biggl(-\frac{1}{2}
\sum_{i,j=1}^{k}
\int\ d^{d}x\int d^{d}y
\,\varphi_{i}(x)
D_{ij}(x-y)\varphi_{j}(x)+\sum_{i=1}^{k}\varphi_{i}J_{i}
\biggr)},
\end{equation}
where we have that
\begin{equation}
\sum_{i=1}^{k}\varphi_{i}J_{i}=\sum_{i=1}^{k}\int d^{d}x\,\varphi_{i}(x)J_{i}(x).
\label{a111}
\end{equation}
Performing the gaussian integrals we get
\begin{equation}
Z_{0\,k}(J_{i})
=\exp\biggl(\frac{1}{2}\sum_{i,j=1}^{k}
\int\,d^{d}x\int\,d^{d}y\,J_{i}(x)\bigl[G_{0}\bigr]_{ij}(m_{0},\sigma; x-y)J_{j}(y)\biggr),
\label{aa125}
\end{equation}
where $\bigl[G_{0}\bigr]_{ij}(m_{0},\sigma; x-y)$ is the two-point correlation function for the replica field theory. We have
\begin{equation}
\sum_{j=1}^{k}\int d^{d}z\,\bigl[G_{0}\bigr]_{ij}(m_{0},\sigma;x-z)\,
D_{jl}(z-y)
=\delta_{il}\delta^{d}(x-y).
\label{zz47}
\end{equation}
We can use $Z_{0\,k}(J_{i})$ to obtain $Z_{k}$ by functional derivatives. The $Z_{k}$ will appear in the definition
of the average free energy. We have
\begin{equation}
Z_{k}
=\Biggl[\exp\biggl(-S_{I}\biggl(\frac{\delta}{\delta\,J_{i}}\biggr)\Biggr)Z_{0\,k}(J_{i})\Biggr]|_{J_{i}=0}.
\label{aa126}
\end{equation}

It is interesting
to analyze the effective action in momentum space. Performing a Fourier transform  we get
\begin{equation}
S_{eff}(\varphi_{i},k)=\frac{1}{2}\sum_{i,j=1}^{k}
\int\,\frac{d^{d}p}{(2\pi)^{d}}\,\,\varphi_{i}(p)\bigl[G_{0}
\bigr]_{ij}^{-1}\varphi_{j}(-p)+\frac{\lambda}{4!}\sum_{i=1}^{k}\varphi_{i}^{4},
\label{aa13}
\end{equation}
where, in the quadratic part of $S_{eff}(\varphi_{i},k)$, the quantity $\bigl[G_{0}\bigr]_{ij}^{-1}$ is the inverse of the
two-point correlation function.  In this tree-level approximation we have
\begin{equation}
\bigl[G_{0}\bigr]_{ij}^{-1}(p)=(p^{2}+m_{0}^{2})\delta_{ij}-\sigma.
\label{aa14}
\end{equation}
To invert this matrix, let us express $\bigl[G_{0}\bigr]_{ij}^{-1}$ in terms of
appropriate projector operators,
\begin{equation}
\bigl[G_{0}\bigr]_{ij}^{-1}(p)=(p^{2}+m_{0}^{2})\biggl(\delta_{ij}-\frac{1}{k}\biggr)+(p^{2}+m_{0}^{2}-k\sigma)\frac{1}{k}.
\label{aa15}
\end{equation}
The projectors operators $(P_{T})_{ij}$ and $(P_{L})_{ij}$ are defined, respectively, as
\begin{equation}
(P_{T})_{ij}=\delta_{ij}-\frac{1}{k}
\label{aa16}
\end{equation}
and
\begin{equation}
(P_{L})_{ij}=\frac{1}{k}.
\label{aa17}
\end{equation}
Using the projectors operators we can write the
two-point correlation function $\bigl[G_{0}\bigr]_{ij}(p)$ as
\begin{equation}
\bigl[G_{0}\bigr]_{ij}(p)=\frac{\delta_{ij}}{(p^{2}+m_{0}^{2})}+\frac{\sigma}{(p^{2}+m_{0}^{2})(p^{2}+m_{0}^{2}-k\sigma)}.
\label{aa18}
\end{equation}
The first term in the right hand side of Eq. (\ref{aa18}) is the bare contribution to the connected two-point correlation
function; the second term is the contribution to the disconnected two-point correlation
function, which becomes connected after averaging the disorder. The first term is characteristic of pure systems
and the additional term, which becomes a squared Lorentzian term (for $k\rightarrow\,0$),
appears in these random field systems \cite{rfs,rfs2}.

In the next section we present a mathematical rigorous technique for computing the average free
energy of a disordered system with quenched randomness using the integer moments of
the partition function, without taking the limit of these integers to zero. The basic tool of
this method is a complex function that we call the distributional zeta-function.

\section{The distributional zeta-function in disordered systems}

\quad

The purpose of this section is to present an alternative method to
calculate the average free energy of a system with quenched disorder, using the replicas.
Let ${\cal M}$ be a Euclidean manifold. The fundamental objects here are the fields, a
space $C^{\infty}({\cal M},\mathbb{R})$ of smooth functions on ${\cal M}$.
Let $S:C^{\infty}({\cal M},\mathbb{R})\rightarrow\,\mathbb{R}$ be an action functional.
%
The Euclidean action functional associated with the model with disorder degrees of freedom can be written
as
\begin{equation}
S(h,\varphi)=\int d^{d}x\, \biggl(\frac{1}{2}
\varphi(x)\Bigl(-\Delta\,+m_{0}^{2}\Bigr)
\varphi(x)
+\frac{\lambda}{4!}\varphi^{4}(x)-h(x)\varphi(x)\biggr).
\label{dis1}
\end{equation}
Again, the symbol $\Delta$ denotes the Laplacian in $\mathbb{R}^{d}$.
The $h$-dependent free energy $F(h)$ is given by
\begin{equation}
F(h)=\ln\, \int[d\varphi]  \exp\left(-S+ \int d^{d}x\,
h(x)\varphi(x)\right),
\label{fe}
\end{equation}
where, as above specified,  $S$ is the action including $S_0$ and $S_I$, i.e., the free and interaction contributions.
The average free energy $F_{q}$ is defined as
\begin{equation}
F_{q}=\int\,d\rho[h]\,F(h).
\label{sa27}
\end{equation}
Recall that  $h$ is a random variable with probability distribution $d\rho[h]=d[h] P(h)$
with $d[h]=\prod_{x} dh(x)$.
Direct use of these definitions yields
\begin{equation}
F_{q}=\int\,d[h]P(h)\ln Z(h).
\label{sa2}
\end{equation}

On way
to compute the average free energy is to use the standard replica method, briefly discussed in the previous section.
Here we propose an alternative procedure to find the average free energy
by means of a \emph{distributional zeta-function}.
Observe that if  $(X,\mathcal{A},\mu)$ is a measure space and $f:X\to(0,\infty)$ is
measurable, one can define a generalized $\zeta$-function
\begin{equation}
\zeta_{\,\mu,f}(s)=\int_X f(x)^{-s}\, d\mu(x)
\end{equation}
for those $s\in\mathbb{C}$ such that  $f^{-s}\in L^1(\mu)$,
where in the above integral $f^{-s}=\exp(-s\log(f))$ is obtained using the
principal branch of the logarithm.
%
This formalism encompasses some well-known instances of zeta-functions for $f(x)=x$:
if $X=\mathbb{N}$ and $\mu$ the counting measure we retrieve the
Riemann zeta-function \cite{riem,riem2}; instead, if $\mu$ counts only the prime numbers, we
retrieve the prime zeta-function \cite{landau,fro}
finally, if $X=\R$ and $\mu$ counts the eigenvalues of an elliptic operator, with
their respective multiplicity, we obtain the spectral zeta-function \cite{hawking}.
Further extending this formalism to the case where $f(h)=Z(h)$ and $d\mu(h)=d[h]P(h)$ leads
to the definition of the distributional zeta-function $\Phi(s)$,
\begin{equation}
\Phi(s)=\int d[h]P(h)\frac{1}{Z(h)^{s}}
\label{pro1}
\end{equation}
for $s\in \mathbb{C}$, this function being defined in the region where the
above integral converges.
We will reproduce the steps of the use of the spectral zeta-function
for computing functional determinants in field theory,
 proving that
$\Phi(s)$ is well defined for $\real(s) \geq 0$, that $F_q=-(d/ds)\Phi(s)|_{s=0}$ and
using this equality for computing a series approximation
for the average free energy $F_q$.


Let us prove that $\Phi(s)$ is defined for $\real(s) \geq 0$. Since $S_0$
and $S_I$ given by Eqs. (\ref{9}) and (\ref{10}) are even (as functions of $\varphi$), $Z(h)=Z(-h)=(Z(h)+Z(-h))/2$
and
\begin{equation}
  Z(h)=\int [d\varphi]\,\, \exp\left(-S\right) \;
  \cosh\left( \int d^{d}x\,
h(x)\varphi(x)\right).
\label{8b}
\end{equation}
Hence $Z(h) \geq Z(0)$ and, for $\real(s) \geq 0$, we have
\begin{equation}
  \label{eq:8c}
  \int d[h]P(h)\left|\frac{1}{Z(h)^{s}}\right|\leq\int d[h]P(h)
  \frac{1}{Z(0)^{\real(s)}}=
   \dfrac{1}{Z(0)^{\real(s)}}<\infty.
\end{equation}
Therefore, the integral in Eq.\ \eqref{pro1} is well defined in the half
complex plane $\real(s)\geq 0$
and $\Phi$ is also defined in this region, \emph{without} resorting to analytic continuations.

Since $-(d/ds)Z(h)^{-s}|_{s=0^{\,+}}=\ln Z(h)$\footnote{$(d/ds)f|_{s=0^+}$ stands
for $\lim_{s\to 0^+}\frac{f(s)-f(0)}{s}$, whenever this limit exists}, we can write
\begin{align}
F_{q}=-\int d[h]P(h)\frac{d}{ds}\frac{1}{Z(h)^{s}}|_{s=0^{\,+}}
=-\frac{d}{ds}\Phi(s)|_{s=0^{\,+}},
\label{sa22}
\end{align}
where the second equality is justified by the fact that $Z(h)\geq Z(0)$
and an application of Lebesgue's dominated convergence theorem, if we interpret
$d[h]P(h)$ as a measure. Observe that we obtained an analytic expression
for  $F_q$ which, contrary to the standard replica method, \emph{does not involve derivation
of the (integer) moments of the partition function}.
Direct use of Euler's integral representation for the gamma function give us
\begin{equation}
\frac{1}{Z(h)^{s}}=\frac{1}{\Gamma(s)}
\int_{0}^{\infty}dt\,t^{s-1}e^{-Z(h)t},\,\,\,\,\, \text{for}\,\,\,\, \real\,(s)>0.
\label{me5}
\end{equation}
Although the above Mellin integral converges only for $\real(s)>0$, as  $Z(h)>0$, we will show how
to obtain from the above expression a formula for the free energy valid for $\real\,(s) \geq 0$.
Substituting Eq.\ \eqref{me5} in Eq.\ \eqref{pro1} we get
\begin{equation}
\Phi(s)=\dfrac{1}{\Gamma(s)}\int d[h]P(h)\int_0^\infty dt\,t^{s-1}e^{-Z(h)t}.
\label{pro1.b}
\end{equation}
We already know that the distributional zeta function $\Phi(s)$ is defined for $\mathrm{Re}(s) \geq 0$.
Now we will use the above expression for computing its derivative at $s=0$
by analytic tools.
We assume at principle the commutativity
of the following operations, disorder average, differentiation, integration if necessary.

From the above discussion, it follows that the average free energy $F_{q}$
can be written as
\begin{equation}
F_{q}=-\frac{d}{ds}\frac{1}{\Gamma(s)}\int d[h]P(h)\int_{0}^{\infty}\,dt\,t^{s-1} e^{-Z(h)t}|_{s=0^{\,+}}.
\label{m22}
\end{equation}
To continue, take $a>0$ and write $\Phi=\Phi_1+\Phi_2$ where
\begin{equation}
\Phi_{1}(s)=\frac{1}{\Gamma(s)}\int d[h]P(h)\int_{0}^{a}\,dt\, t^{s-1}e^{-Z(h)t}
\label{m23}
\end{equation}
and
\begin{equation}
\Phi_{2}(s)=\frac{1}{\Gamma(s)}\int d[h]P(h)\int_{a}^{\infty}\,dt\, t^{s-1}e^{-Z(h)t}.
\label{m24a}
\end{equation}
The average free energy can be written as
\begin{equation}
F_q=-\frac{d}{ds}\Phi_1(s)|_{s=0^{\,+}}-\frac{d}{ds}\Phi_2(s)|_{s=0}\,.
\end{equation}

The integral $\Phi_{2}(s)$ defines an analytic function defined in the whole complex plane.
In the innermost integral in $\Phi_{1}(s)$ the
series representation for the exponential converges uniformly (for each $h$), so that
we can reverse the order on integration and summation to obtain
\begin{equation}
\Phi_{1}(s)=\int d[h]P(h)\frac{1}{\Gamma(s)}\sum_{k=0}^\infty
\frac{(-1)^{k}a^{k+s}}{k!(k+s)}Z(h)^k.
\label{m23b}
\end{equation}
The term $k=0$ in the above summations contain a removable singularity at $s=0$ because $\Gamma(s)s=\Gamma(s+1)$,
so that we can write
\begin{equation}
\Phi_{1}(s)=\frac{a^s}{\Gamma(s+1)}+\frac{1}{\Gamma(s)}\sum_{k=1}^\infty \frac{(-1)^ka^{k+s}}{k!(k+s)}\,
\mathbb{E}\,{Z^{\,k}},
\label{m23c}
\end{equation}
an expression valid for $\real(s)\geq\,0$.
The function $\Gamma(s)$ has a pole in at $s=0$ with residue $1$, therefore
\begin{equation}
-\dfrac{d}{ds}\Phi_{1}(s)|_{s=0^{\,+}}=\sum_{k=1}^\infty \frac{(-1)^{k+1}a^{k}}{k!k}\,
\mathbb{E}\,{Z^{\,k}}
+f(a),
\label{m23d}
\end{equation}
where
\begin{equation}
f(a)=-\dfrac{d}{ds}\left(\dfrac{a^s}{\Gamma(s+1)}\right)|_{s=0}
=-\bigl(\ln(a)+\gamma\bigr)
\label{m23e2}
\end{equation}
and $\gamma$ is Euler's constant $ 0.577\dots$
We have just shown that the contribution of $-(d/ds)\Phi_1(s)|_{s=0^{\,+}}$
to the average free energy (see Eq. (\ref{m23d})) can be written as a series in which all integer moments of the partition
function appear. Although such a representation cannot be obtained, in general, for the contribution
coming from $-(d/ds)\Phi_2(s)|_{s=0}$, we will show how to bound this contribution.
Using again the fact that $\Gamma(s)$ has a pole at $s=0$ with residue $1$ and
taking the derivative of the product
that defines $\Phi_2$ in Eq. (\ref{m24a}), we conclude that
\begin{equation}
-\dfrac{d}{ds}\Phi_{2}(s)|_{s=0}=-\int d[h]P(h)\int_{a}^{\infty}\,\dfrac{dt}{t}\, e^{-Z(h)t}
=R(a).
\label{m24}
\end{equation}
Therefore, using again the inequality $Z(h)\geq Z(0)$ we obtain the bound
\begin{equation}
\left|-\dfrac{d}{ds}\Phi_{2}(s)|_{s=0}\right| \leq
\int d[h]P(h)\int_{a}^{\infty}\,\dfrac{dt}{t}\, e^{-Z(0)t}
\leq\dfrac{1}{Z(0)a}\exp\big(-Z(0)a\big).
\label{m24.b}
\end{equation}
In conclusion, using the distributional zeta-function we are able to
represent the average free energy as
\begin{equation}
F_q=\sum_{k=1}^\infty \frac{(-1)^{k+1}a^{k}}{k!k}\,
\mathbb{E}\,{Z^{\,k}}
-\bigr(\ln(a)+\gamma\bigl)+R(a),
\qquad \left|R(a)\right|\leq \dfrac{1}{Z(0)a}\exp\big(-Z(0)a\big).
\label{m23e1}
\end{equation}
Observe that we cannot take the limit $a\to\infty$, because in this case the
above series become meaningless.
However, the contribution of $R(a)$ to the free energy can be made
as small as desired, taking $a$ large enough.
A case of special interest is $a=1$, because with this choice
the average free energy
can be written as
\begin{equation}
F_q=\sum_{k=1}^\infty \frac{(-1)^{k+1}}{k!k}\,
\mathbb{E}\,{Z^{\,k}}
-\gamma+R(1),
\qquad \left|R(1)\right|\leq \dfrac{1}{Z(0)}\exp\big(-Z(0)\big).
\label{m23f}
\end{equation}
Defining a distributional zeta-function and using analytic tools,
we showed that all replicas contribute to  the average value of the free energy of the system.
Using Eq. (\ref{aa126}) and Eq. (\ref{m23f}) we can write the average free energy in terms of the vacuum to vacuum diagrams.


The main result of this paper is Eq. (\ref{m23f}). Instead of the limit to zero replicas as in the usual
procedure, we obtained an expression for the average free energy  as a series where all the replicas contribute
plus another term which can not be
represented by a series on the replicas partition function.
For $Z(0)$ bounded away from zero, the second term contribution's can be made as small as desired.
Note that the representation of the average free energy just by a series on the the replicas partition functions
would not describe the loss of analyticity on the average free energy in the $Z(h)\to 0$ phase transitions.

Two comments are in order. First,
in those phase transitions where $Z(h)\to 0$, the divergence of the free energy comes from $R(a)=-(d/ds)\Phi_2(s)$
as revealed by direct inspection of Eq. (\ref{m24}); the contribution to the free energy due to the series expansion
captures its non-analytic behaviour when $Z(h)\to\infty$.
Second, in the
replica field theory the symmetry group is $S_{k}$ (the permutation group of $k$ elements). One advantage of the distributional zeta-function technique
is that it is not necessary to understand the properties of the permutation group when $k\,\rightarrow\,0$. 
Let us discuss the replica symmetric ansatz, i.e., suppose that $\varphi_{i}(x)=\varphi(x)$. For the case of identical
replicas, the saddle-point equations  make sense for $k<\frac{m_{0}^{2}}{\sigma}$. Therefore, in this case one could justifies this replica
symmetric solution. For $k\geq\frac{m_{0}^{2}}{\sigma}$, we must define a shifted field $\phi(x)=\varphi(x)-v$, for
$v=(\frac{k\sigma-m_{0}^{2}}{6\lambda})^{1/2}$. In terms of this new field $\phi(x)$, we get a positive mass squared with new self-interaction vertices
$\phi^{3}(x)$ and $\phi(x)^{4}$. Therefore the symmetry between the replicas
is broken using the results of the saddle-point equations of the model. 

\section{Conclusions}

\quad

There is a growing interest in disordered systems in physics and many areas beyond physics.
In statistical mechanics, the random magnetic field Ising model has been investigated in the last forty years.
The hamiltonian of the random field Ising model is analogous to the one of the
classical Ising model, but allowing for a quenched random magnetic field.
This model can be analyzed using a collective description instead of
dealing with individual elements. The hamiltonian of the discrete model can be replaced by an effective hamiltonian of the
Landau-Ginzburg model, where the order parameter is a continuous field $\varphi(x)$ coupled to
a quenched random field. The usual approach to study such disordered systems is to transform the random problem into a
translational invariant one. For quenched disorder, one is mainly interested in
averaging the free energy over the disorder, which amounts to averaging the log of the partition function $Z$,
the connected vacuum to vacuum diagrams.
The replica method is a powerful tool used to calculate the free energy of systems with quenched disorder.
Despite the absence of a mathematically rigorous derivation, the standard replica method provides correct results in many situations.
As we discussed, it is  natural to ask if it
is possible to find a mathematically rigorous derivation which legitimates the use of the replica partition functions for
computing the average free energy of the system.
In this paper we present a rigorous derivation of the average free energy in the continuous
version of the random field Ising model, as the sum
of a series using the replica partition functions plus a contribution which, far from
phase transitions, can be made as small as desired.

In order to find the average free energy we define a distributional zeta-function.
The derivative of the distributional zeta-function at $s=0$ yields the average free energy.
Making use of the Mellin transform
and analytic continuation,
it is possible to obtain a
series representation for the average free energy where all the
integer moments of the partition function of the model contribute.
Our method is closely related to the use of spectral zeta-functions for
evaluating the determinant of elliptic differential operators in Euclidean field
theory. This technique also has been used to calculate the renormalized vacuum energy
of quantum field in the presence of boundaries.

The distributional zeta-function technique is a mathematically rigorous method for computing
the average free energy using the replica partition functions. Further validation of the distributional zeta-function technique in the
Sherrington-Kirkpatrick \cite{sk} and also the random energy model 
is the natural continuation of the present work.
In the random energy model
it is known that for $k<1$ there is a phase transition at some critical temperature, but
high and low temperature regimes exchange their roles.
Also in the Sherrington-Kirkpatrick model, after integrating over the
random spin couplings and using the saddle-point approximation, the average free energy density
is written in terms of an order parameter $k\times k$ matrix $Q_{ij}$.
The stability of the model implies that the Hessian matrix $\frac{\partial^{2}F}{\partial Q_{ij}\partial Q_{kl}}$, $H$ has positive
eigenvalues. For example, in the subspace of the matrix satisfying
$Q_{ij}=q$, the condition above leads to $\frac{1}{k} Tr Q^{2}=(k-1)q^{2}$, which is always valid for $k>1$.
In the case of the Sherrington-Kirkpatrick model, at some stage of the calculations we must use the
saddle-point equations to extremize the free energy with respect to some variable. As stressed by Dotsenko in Ref.\cite{sum}, it is
not clear that this condition of extremizing the free energy can be used in this formalism, where all the moments of the
partition function appear in the average free energy.

There are many papers studying finite size-effects in Euclidean field theory \cite{finitez1,finitez2,finitez3,finitez4,khanna}.
Suppose a Euclidean field theory with a
$(\lambda\varphi^{4})$ self-interaction defined in a
$d$-dimensional Euclidean space. Assume that the domain where the field
is defined is $S^{1}\times \mathbb{R}^{d-1}$. For very small radius, since there is a size-dependent renormalized mass square,
the system is in the disordered phase.
Increasing the radius of the compactified dimension, we
obtain the critical size where a second order phase transition occurs \cite{ford, adolfo1}.
A complete discussion of finite size effects in phase transitions can be found in \cite{danchev}.
For the reader interested in the study of finite size effects in random systems, see Refs. \cite{s1,s2,ricci1,ricci2}.

Recently, it was discussed a phase transition in the
finite-size disordered $\lambda\varphi^{4}$ model, using the standard replica method.
The authors considered  the scalar model
system at zero temperature defined in a space with periodic boundary conditions in one space dimension.
It was shown that there is a critical length
where the system develop a second-order phase transition, where the system presents long-range correlations with
power law decay \cite{periodic}.
As an application of
our approach, it would be interesting to discuss finite-size effects in one disordered $\lambda\varphi^{4}$ model in
a $d$-dimensional Euclidean space using this distributional zeta-function technique. Does
the renormalized mass obtained from the distributional zeta-function coincide
with the renormalized mass obtained by the usual replica method where the average free energy
is obtained from a zero component field theory? This problem also will be subject of future investigations.

\section{Acknowledgments}
We would like to thanks G. Parisi and S. Fulling for encouragements.
We would like to acknowledge also V. Dotsenko, G. Krein, E. Curado, T. Micklitz, S. Alves Dias, F. Nobre and
G. Menezes for the fruitful discussions.
This paper was supported by Conselho Nacional de
Desenvolvimento Cientifico e Tecnol{\'o}gico do Brazil (CNPq).

\begin{appendix}
\makeatletter \@addtoreset{equation}{section} \makeatother
\renewcommand{\theequation}{\thesection.\arabic{equation}}

\section{The spectral zeta function in Euclidean field theory}

Here we present a new mathematical rigorous technique for computing the average free
energy of a disordered system with quenched randomness, using the replicas.
The basic tool of this technique is
a distributional zeta-function.
In this appendix  we would like to discuss the similarities of our approach with the spectral zeta-function regularization
in Euclidean field theory \cite{dowker,hawking}.

If $\lambda_{k}$ is a sequence of non-zero complex numbers, we define the zeta regularized
product of these numbers as
\begin{equation}
\prod_{k}\,\lambda_{k}=e^{-\frac{d}{ds}\,\zeta(s)|_{s=0}},
\label{me7}
\end{equation}
where
\begin{equation}
\zeta(s)=\sum_{k}\,\frac{1}{\lambda_{k}^{s}}\,\,\,  for\,\,\,\, Re(s_{0})>0,
\label{me8}
\end{equation}
is the zeta-function associated with the sequence $\lambda_{k}$.
If $\lambda_{k}$ is the sequence of the
positive eigenvalues of the Laplacian on a manifold, then the zeta regularized product is the
determinant of the Laplacian. The free energy of a system with this eigenvalues in this case is
\begin{equation}
F=\frac{1}{2}\frac{d}{ds}\,\zeta_{D}(s)|_{s=0}.
\label{me9}
\end{equation}
In order to obtain the free energy of a system, we have to show that $\zeta_{D}(s)$ has a meromorphic continuation
with at most simple poles, to the half-plane containing the origin and analytic at the origin.
We have scaling properties, i.e.,
\begin{equation}
-\frac{d}{ds}\,\zeta_{\mu D}(s)|_{s=0}=-\frac{d}{ds}\,\zeta_{D}(s)|_{s=0}+\ln\mu\,\zeta_{D}(s)|_{s=0}.
\label{sca}
\end{equation}
Note that
there are sequences of numbers which are not zeta regularized, where the free energy is not defined \cite{primes0,primes}.
Suppose that the sequence of numbers is the sequence of the prime numbers. In this case the prime
zeta function $P(s)$, $s=\sigma+i\tau$, for $\sigma,\tau\,\in \mathbb{R}$ is defined as
\begin{equation}
P(s)=\sum_{p}p^{-s}, \,\,\,\,Re(s)>1.
\label{me10}
\end{equation}
Introducing the M\"obius function $\mu(n)$ \cite{hardy} we can write
\begin{equation}
P(s)=\sum_{k}\frac{\mu(k)}{k}\,\ln \zeta(ks),
\label{me11}
\end{equation}
where $\zeta(s)$ is the Riemann zeta function \cite{riem}. The line $Re(s)=0$ is a natural boundary of $P(s)$. Therefore the prime
zeta function can be analytically extended only in the strip $0<\sigma\leq 1$ \cite{landau,fro}.
Since we need to compute the derivative of the spectral zeta function at $s=0$, the free energy is not defined.
The sequence of primes numbers is not zeta-regularized.

In order to discuss the similarities of our approach with the spectral zeta-function regularization
in Euclidean field theory, let us discuss the heat-kernel on Euclidean manifolds. Let 
${\cal M}$ be an arbitrary smooth connected $d$-dimensional compact Riemannian manifold. The
Riemannian structure allow to introduce on ${\cal M}$ volumes of all dimensions. The
Riemannian $d$-volume is defined as $d\mu={\sqrt{g}}dx^{1}...dx^{d}$.
Now consider the following initial boundary problem in $(0,\infty)\times \mathbb{R}^{d}$.
We have
\begin{equation}
\frac{\partial}{\partial t}u(t,x)=\Delta\,u(t,x),
\label{me12}
\end{equation}
with the initial condition
\begin{equation}
u(0,x)=f(x),
\label{me13}
\end{equation}
and also
\begin{equation}
u(t,x)|_{x\in\,\partial\,\Omega}=0,
\label{me14}
\end{equation}
where the symbol $\Delta$ denotes the Laplacian in $\mathbb{R}^{d}$.
The evolution equation has the solution
\begin{equation}
u(t,x)=e^{t\Delta_{\Omega}}f(x).
\label{me15}
\end{equation}
Suppose that the function $f(x)$ is $f\,\in L^{2}(\Omega)$.
The standard technique of the spectral theory of elliptic operators
implies that there exists a orthonormal basis $[\phi_{k}]$
in $L^{2}(\Omega)$ such that $\phi_{k}$ is a eigenfunction of $-\Delta$ in
$\Omega$ with the eigenvalue $\lambda_{k}=\lambda_{k}(\Omega)$, and
$0\leq\lambda_{1}\leq\lambda_{2}...\lambda_{k}\rightarrow \infty$, when $k\rightarrow\infty$.
Therefore we can write the
following series representation
\begin{equation}
f(x)=\sum_{k=1}^{\infty}a_{k}\,\phi_{k}(x),
\label{me16}
\end{equation}
where
\begin{equation}
a_{k}=\int_{\Omega}\,d\mu(y)f(y)\phi_{k}(y).
\label{me17}
\end{equation}
We can write
\begin{equation}
u(t,x)=\int_{\Omega}\,p_{\Omega}(t,x,y)f(y)d\mu(y),
\label{me18}
\end{equation}
where
\begin{equation}
p_{\Omega}(t,x,y)=\sum_{k=1}^{\infty}\,e^{-\lambda_{k}(\Omega)}\phi_{k}(x)\phi_{k}(y).
\label{me19}
\end{equation}
We denote $p_{\Omega}(t,x,y)$ the heat-kernel of $\Omega$. The operator $e^{t\Delta_{\Omega}}$ has the integral kernel
$p_{\Omega}(t,x,y)$. We have
\begin{equation}
\int_{\Omega}d\mu(y)\,p_{\Omega}(t,y,y)=\sum_{k=1}^{\infty}\,e^{-t\lambda_{k}(\Omega)}
\label{me20}
\end{equation}
Recall that the Mellin transform
of $f:\,\mathbb{R}_+\rightarrow \mathbb{C}$ is defined as
\begin{equation}
M[f;s]=\int_{0}^{\infty} \frac{dx}{x}f(x)x^{s}
\label{me1}
\end{equation}
for those $s\,\in \mathbb{C}$ such that
$f(x)\,x^{s}\,\in \,L^{1}(\mathbb{R}_+)$.
For example, if $f(x)=e^{-x}$ then the Mellin transform is the gamma function, i.e.,
\begin{equation}
\Gamma(s)=\int_{0}^{\infty}dt\,t^{s-1}e^{-t},\,\,\,\,\, for\,\,\,\, Re(s)>0.
\label{me2}
\end{equation}

We can write that the free energy is given by
\begin{equation}
F=\frac{1}{2}\frac{d}{ds}\Biggl[\frac{M[\int d\mu(y)p_{\Omega}(t,y,y),s]}{M[e^{-t},s]}\Biggr]|_{s=0}.
\label{m21}
\end{equation}
Our mathematical approach to make rigorous the replica method using a distributional zeta-function can be compared with the
spectral zeta-function method widely used in quantum field theory. Using the Eq. (\ref{m22}) the average
free energy of the continuous version for the random field Ising model can be written as
\begin{equation}
F_{q}=-\frac{d}{ds}\Biggl[\frac{M[\int d[h]P(h)e^{-Z(h)t},s]}{M[e^{-t},s]}\Biggr]|_{s=0^+}.
\label{me6}
\end{equation}
Note the similarities between the Eq. (\ref{m21}) and Eq. (\ref{me6}).

\end{appendix}

\end{document}